\documentclass[preprint,superscriptaddress,double-spaced,floatfix,longbibliography]{revtex4-1}
\usepackage{multirow}
\usepackage{graphicx,amsmath,bbm,bm,array,subfigure}
\usepackage{appendix}
\usepackage{lipsum}
\usepackage{dsfont}
\usepackage{calrsfs}
\usepackage{lineno}   
\usepackage{bbold}
\usepackage[linktocpage=true,colorlinks=true,linkcolor=blue,citecolor=blue,urlcolor=blue]{hyperref}
\usepackage{MnSymbol}
\newcommand\Snn{\mathcal{S}_{\rho\rho}(\vec{k},\omega)}

\begin{document}
\title {Navier-Stokes hydrodynamics near the critical point with out-of-equilibrium modes}
\author{Md Hasanujjaman}
\email{hasan@apcrgc.org}
\affiliation{Department of Physics, A. P. C. Roy Government College, Siliguri-734010, India}

\author{Golam Sarwar}
\email{golamsarwar1990@gmail.com}
\affiliation{Independent Researcher, Kent, England, United Kingdom}
\author{Mahfuzur Rahaman}
\email{mahfuzurrahaman01@gmail.com}
\affiliation{Department of Physics, Maulana Azad College, Kolkata - 700013, India}
\author{Abhijit Bhattacharyya}
\email{abhattacharyyacu@gmail.com}
\affiliation{Department of Physics, University of Calcutta,  Kolkata-700009, India}

\author{Jane Alam}
\email{janephysics1996@gmail.com}
\affiliation{Murshidabad Maharaja Krishnanath University, Berhampore, Murshidabad-742101, India}

\def\zbf#1{{\bf {#1}}}
\def\bfm#1{\mbox{\boldmath $#1$}}
\def\hf{\frac{1}{2}}
\def\sl{\hspace{-0.15cm}/}
\def\omit#1{_{\!\rlap{$\scriptscriptstyle \backslash$}
{\scriptscriptstyle #1}}}
\def\vec#1{\mathchoice
{\mbox{\boldmath $#1$}}
{\mbox{\boldmath $#1$}}
{\mbox{\boldmath $\scriptstyle #1$}}
{\mbox{\boldmath $\scriptscriptstyle #1$}}
}
\def \beq{\begin{equation}}
\def \eeq{\end{equation}}
\def \beqa{\begin{eqnarray}}
\def \eeqa{\end{eqnarray}}
\def \pd{\partial}
\def \nn{\nonumber}
\begin{abstract}
We investigate how slow out-of-equilibrium modes (OEM), introduced to extend the regime of validity of hydrodynamics near the critical point, fundamentally reshape the power spectrum of dynamical density fluctuations. We have used the equation of motion of slow modes for the situation when the extensive nature of thermodynamics is not altered due to the introduction of OEM. We find that the extensivity condition puts an extra constraint  on the coupling of OEM  with the four divergence of velocity. In the absence of OEM, the dynamic structure factor exhibits three Lorentzian peaks, one
at zero frequency ($\omega$), called Rayleigh peak and the other two called Brillouin  peaks located 
symmetrically about $\omega=0$. When the system is away from the critical point the inclusion of the OEM 
preserves the symmetry of the distributions but all the
peaks acquire reduced widths and the Brillouin peaks  adopt unequal heights. 
Irrespective of the value of OEM (zero or nonzero) the Brillouin peaks vanish near the critical point.  
The width of the Rayleigh  peak substantially reduced in presence  of OEM. 
Such reductions of the widths in presence of the OEM indicates the reduction in the decay rate of the fluctuation 
which leads to slowing down of decay of the fluctuations, a distinctive feature of the critical point. 
\end{abstract}
\maketitle

\section{Introduction}
\label{sec1}
The fluid dynamics provides long wavelength behaviour of system in local thermal  equilibrium. A system 
can be considered in thermal equilibrium and therefore, the hydrodynamics can be applied 
if the Knudsen number, $K_n<<1$, where $K_n=\ell/R$, $\ell$ is
the mean free path and $R$ is the characteristic system size. 
Stated another way the fluid dynamical representation is permitted in the regime $k\,\xi<<1$~\cite{Stanley}, 
where $k$ and $\xi$ are respectively the wave number and the correlation length of the fluid. Expressing the 
wave number in terms of the wavelength ($\lambda$), the criteria becomes $\xi<<\lambda$ for the 
applicability of the fluid dynamics. This criterion is known to be valid in the relativistic as well as 
in the non-relativistic domain.

We know that different orders of hydrodynamic equations are obtained by expanding the energy-momentum tensor (EMT) in 
powers of gradient of the hydrodynamic variables. Truncating the expansion at the lowest (zeroth) order gives rise to the theory of ideal fluid dynamics. 
Truncation at first and second order in the gradient of the hydrodynamic fields give rise to first-order and second-order hydrodynamic theories, respectively. 
The first-order dissipative hydrodynamics described by the Navier-Stokes (NS)~\cite{Navier1,Navier2,Stokes} equations is known to violate the causality and gives numerically 
unstable solutions hence it becomes unfit in relativistic domain~\cite{Hiscock:1983zz,Hiscock:1985zz}. Therefore, the second-order theory {\it{i.e.,}} the M\"uller-Israel-Stewart (MIS) ~~\cite{Muller:1967zza,Israel:1979wp} theory which preserves causality and provides stable solutions is used to describe relativistic viscous fluid. Despite its great success, 
such theories can not be applied to describe the fluid away from equilibrium~\cite{Romatschke:2017vte,Romatschke:2017ejr}.

The NS equation is routinely used to describe fluid flow in non-relativistic domain. However, the validity of hydrodynamical description turns out to be incapacitated near the critical point (CP) of fluid because the condition $\xi<<\lambda$ mentioned above is not satisfied as the correlation length diverges near CP~~\cite{Stanley,Hohenberg}. 
Therefore, NS equation can not be applied near CP even in the non-relativistic domain.  
The fluid falls out of equilibrium due to the divergence of the correlation length near CP, as a consequence, it becomes meaningless to study the prediction of hydrodynamics near the CP. However, the endeavour of studying the critical point within the hydrodynamic framework does not end here. The schemes to deal with such an 
out-of-equilibrium fluid are discussed in the Refs.~\cite{Romatschke:2017vte,Romatschke:2017ejr,Stephanov:2017ghc}. 
In Refs.~\cite{Romatschke:2017vte,Romatschke:2017ejr}, authors proposed that the hydrodynamics can still be applied to 
fluids that are far away from equilibrium by taking the higher-order gradients of the hydrodynamic fields into the EMT. 
Nevertheless, the gradient expansion scheme is not always convergent~\cite{Heller} and it fails to explain 
the fluid with large fluctuation, which is expected near the critical point. 
Therefore, we have not applied the gradient expansion scheme in the present study. The other possible way of extending the validity of hydrodynamics is known as Hydro+ scheme~\cite{Stephanov:2017ghc} where a parametric slow mode, $\phi$, along with the other hydrodynamical variables are incorporated into the definition of entropy to establish the hydrodynamic equations. The presence of $\phi$ and its coupling with different hydrodynamic fields to allow the study large fluctuations near the critical point.

{\textcolor{black}{The Hydro+ approach \cite{Stephanov:2017ghc,Rajagopal:2020} evaluates the $\phi(t, \vec{x}, \vec{k})$ as the correlator of equal-time two-point function of the order parameter field as:
\begin{equation}
\phi(t, \vec{x}, \vec{k}) = \int d^3\vec{y} \, e^{-i\vec{k} \cdot \vec{y}} \,\Big< \delta M(t,\,\vec{x} + \vec{y}/2) \delta M(t,\,\vec{x} - \vec{y}/2) \Big>.
\end{equation}
Here, $\delta M(t,\,x)=M(t,\,x)-<M(t,\,x)>$, represents the fluctuation in order parameter field, and $\vec{k}$ 
is the wave-vector. The function $\phi$ is proportional to the width of the probability distribution of 
the fluctuation in order parameter, $\delta M$~\cite{Rajagopal:2020}.
The choice of $\phi$ depends on the choice of the system, e.g., it is the local magnetization in a magnet, concentration difference in a binary fluid, the density fluctuation in a liquid-gas
critical point, or an amplitude of a soft phonon. One may think of this $\phi$ as the cloudiness of the medium which may trigger the opalescence of the system.}}

In the present work, the role of $\phi$ is taken into account to extend the validity of the non-relativistic NS 
theory to study the large fluctuation near the critical point. In condensed matter physics,  
the correlation of the density fluctuation are extensively studied to understand the system near CP. The time-dependent 
correlation of density fluctuation leads to the dynamic structure factor [$\Snn$] (defined later), 
which is measured by light/neutron scattering in the laboratory. The $\Snn$ typically exhibits a central 
Rayleigh peak accompanied by two symmetric side Brillouin peaks~\cite{Stanley,Kadanoff1963}. 
The $\Snn$ contains a vast amount of information on the thermodynamic response function 
and the transport coefficients of the medium, {\it{e.g.,}} the speed of sound ($c_{s}$), isobaric ($C_{V}$) and isochoric specific heats ($C_{P}$), the transport coefficients such as shear viscosity $(\eta)$, bulk viscosity $(\zeta)$, and the thermal conductivity $(\chi)$~\cite{Stanley,Minami:2009hn}.
Here we would like to study how $\phi$ affects the dynamical structure factor near CP.  

The paper is organized as follows: In the Sec.\ref{sec2}, we discuss the governing equations to calculate the density fluctuation. Then we derive the dynamic structure factor from the linear mode analysis. In Sec.\ref{sec3}, we present our result, and finally summarize in Sec.\ref{sec4}.      

\section{Formalism: The governing equations}
\label{sec2}
In this section, we briefly discuss the formulation of the hydrodynamic equations when 
the parametrically slowly evolving scalar mode $\phi$ is introduced. Inclusion of the $\phi$ mode will change the entropy density of the fluid, and within a short time window, the fluid reaches a partial equilibrium state i.e., 
with entropy state say $s_{+}(\epsilon,\,n,\,\phi)$, can be written from the Gibbs-Duhem relation as:
\beqa
\label{eq1}
ds_{+}=\beta_{+} \,d\epsilon_{+}-{\alpha}_{+}\,dn_{+} -\pi\,d\phi\,,
\eeqa
where $\epsilon_{+}$ and $n_{+}$ are energy density and number density respectively, $\beta_{+}=1/T$ ($T$ is the temperature), 
${\alpha}_{+}=\mu_{+}/T$, ($\mu$ is the chemical potential) in the partial equilibrium state. \textcolor{black}{Where the `+' sign has been introduced to denote the thermodynamic quantities in the Hydro+ scheme.}
The quantity $\pi (\epsilon_{+},\,n_{+},\,\phi)$ is also the generalized chemical potential 
corresponding to the extra variable $\phi$. For the non-relativistic fluid considered in the present work, it is convenient to express the thermodynamic relations in terms of the mass density $\rho$ instead of the number density $n$. The Gibbs-Duhem relation now reads as:
\beqa
\label{eq1}
ds_{+}=\beta_{+} \,d\epsilon_{+}-\tilde{\alpha}_{+}\,d\rho _{+}-\pi\,d\phi\,,
\eeqa
where $\tilde{\alpha}_{+}={\alpha}_{+}/m=\mu_{+}/mT$.

After sometime time, the fluid reaches in the state of equilibrium, implying $\pi(\epsilon,\,\rho,\bar{\phi})=0$, the entropy will become:
\beqa
s(\epsilon,\,\rho)=s_{+}(\epsilon_{+},\,\rho_{+},\bar{\phi})\,.
\eeqa
The Hydro+ formalism provides a generalized set of equations governing the hydrodynamic fields. These equations are constructed upon the well-established Navier-Stokes (NS) theory, which is founded on the three fundamental conservation laws: mass conservation, momentum conservation, and energy conservation. Apart from these conversion laws, an equation governing 
the evolution of $\phi$ is required. 

The mass conservation equation appears as:
\beqa
\label{NS1}
\frac{\pd \rho}{\pd t}+\vec{\nabla}.(\rho \vec{v})&=&0\,,
\eeqa
where $v$ is the velocity of the fluid. 

The momentum equation is expressed as
\beqa
\label{NS2}
\rho \left( \frac{\partial \vec{v}}{\partial t} + \vec{v} \cdot \nabla \vec{v} \right) = -\nabla P + \Big(\zeta+\frac{4}{3}\eta\Big) \nabla^2 \mathbf{v}\,,
\eeqa
where  $P$ is the thermodynamic pressure. The terms $\zeta$ and $\eta$ are the bulk and shear viscosity. 

The remaining equation for the energy conservation is written as:
\beqa
\label{NS3}
\rho\,\,\Big[\frac{\pd h}{\pd t}+\vec{v}.\vec{\nabla}h\Big]=\frac{\pd P}{\pd t}+\vec{v}.\vec{\nabla}P+\chi\,\nabla^{2}T\,,
\eeqa
 where  $h=(\epsilon +P)/\rho$ ($\epsilon$ being the energy density) is the specific enthalpy, and $\chi$ is the thermal conductivity. In the above constitutive relations, the thermodynamic variables are replaced by the `+' sign, which could also modify the viscous coefficients such as $\eta,\,\zeta$, and $\chi$. But for simplicity, we have considered that the change in the viscous coefficients due to the introduction of $\phi$ mode is small, and thus neglected.

The evolution of $\phi$ is such that it relaxes to an equilibrium value ($\bar{\phi}$) as the system aims towards equilibrium. However, the evolution can in general be influenced by the gradients present in the system. To find relevant gradients that may affect the slow mode at the critical point, the following observation can be made. At the critical point which we are concerned with, the volume fluctuation becomes prominent~\cite{LL,Weinberg:1971mx}. The divergence of fluid velocity accounts for such volume fluctuations in the hydrodynamic description~\cite{Kovacevic2016,Tripolt_2014}. It is straightforward to to show from continuity equation that $V^{-1}\frac{dV}{dt}=\vec{\nabla}.\vec{v}$. So it is natural to consider the effect of fluid velocity divergence on the slow mode which is introduced to describe the evolution near such a critical point. As we have assumed that the $\phi$ itself relaxes over time to its own equilibrium state $\bar{\phi}$,  the general form of the evolution equation (a relaxation type) of the slow mode in the non-relativistic form can be written as~\cite{Stephanov:2017ghc}:
\beqa
\label{slow}
\frac{\pd \phi}{\pd t}+\vec{\nabla}\phi.\vec{v}&=&-F_{\phi}-A_{\phi}\,(\vec{\nabla}.\vec{v})\,.
\eeqa
{\color{black}{In Hydro+, the scalar mode $\phi$ is a slow, non-hydrodynamic variable. In equilibrium, the $\phi$ maximizes the entropy functional, $s_{+}(\epsilon_{+},\,\rho_{+},\,\phi)$. Away from equilibrium, the entropy deviates from the 
maximum value, and the system tries to relax back by the conjugate thermodynamic force, $\pi(\epsilon_{+},\,\rho_{+},\,\phi)\equiv \big(\frac{\pd s_{+}}{\pd \phi}\big)_{\epsilon_{+},\,\rho_{+}}$. The restoring dynamics of $\phi$ thus involve $\pi$. The restoring force to bring back $\phi\to \bar{\phi}$ is what is denoted by $F_{\phi}$. Therefore, it must vanish at equilibrium for a given value of $\epsilon$ and $\rho$, it push back the $\phi$ to its equilibrium value $\bar{\phi}$, i.e., when $\phi\rightarrow  \bar{\phi}$, the factors $F_{\phi}(\epsilon_{+},\,\rho_{+},\,\phi)=0$ or $\pi(\epsilon_{+},\,\rho_{+},\,\phi)=0$. The $\phi$ must evolve in such a way that the entropy production is non-negative. Therefore, $F_{\phi}$ should be proportional to $\pi$ with a positive coefficient. For a small deviation of $\phi$, i.e., $\delta \phi=\phi-\bar{\phi}$, the form could be of $F_{\phi}=\Gamma_{\phi}\pi$, where $\Gamma_{\phi}$ is the relaxation rate of the slow mode. The equation then reappears as:
\beqa
\frac{\pd \phi}{\pd t}+\vec{\nabla}\phi.\vec{v}&=&-\Gamma_{\phi}\,\pi-A_{\phi}\,(\vec{\nabla}.\vec{v})\,.
\eeqa
The first term on the RHS thus signifies that it drives back the $\phi$ to its equilibrium value. The second term describes how expansion shifts the equilibrium value itself.}}

 The susceptibility of $\phi$ by compression or expansion is given by $A_{\phi}$. In Eq.\eqref{slow}, $F_{\phi}$ and $A_{\phi}$ are the unknown quantities, which could be found out by imposing the second law of thermodynamics to close the equation. The second law of thermodynamics constrains the basic equations. For example, in ordinary hydrodynamics, it demands:  $s=\beta (\epsilon+P)-\tilde{\alpha}\,\rho$, along with the positive coefficients of shear and bulk viscosity, and thermal conductivity. Similarly, in the context of Hydro+ formalism, the second law will restrict the possible forms that $F_{\phi}$ and $A_{\phi}$ can take.
 The entropy four-current is written as~\cite{Eckart:1940te}:
 \beqa
 \label{eq7}
 S_{+}^{\mu}\equiv S_{+}^{\mu}({s}_{+},\,\vec{J}_{s}^{+})\,.
 \eeqa
 where $\vec{J}_{s}^{+}=s_{+}\vec{v}$, the the entropy current vector. The $\vec{J}_{s}^{+}$ gets contribution from the gradient of the hydrodynamic fields, and thus it appears as:
 \beqa
 \vec{J}_{s}^{+}=s_{+}\vec{v}+\Delta \vec{J}_{s}^{+}\,.
 \eeqa
 
  In Appendix-\ref{apA}, we  have shown that,
\beqa
\label{eq8}
s_{+}=\beta_{+}(\epsilon_{+}+P_{+})-\tilde{\alpha}_{+} \rho_{+}-\pi \phi\,,
\eeqa
Using Eq.~\eqref{eq1} and Eq.~\eqref{eq8} we obtain, 
\beqa
\beta_{+} dP_{+}=-(\epsilon_{+}+P_{+})d\beta_{+}+\rho_{+} d\tilde{\alpha}_{+}+\phi d\pi\,.
\eeqa
The  second law of thermodynamics can be achieved by taking the divergence of the entropy current as:
\beqa
\pd_{\mu}S^{\mu}_{+}\ge0\,.
\eeqa
Expanding we get
\beqa
\pd_{t}s_{+}+\vec{\nabla}.\vec{J}_{s}^{+}\ge 0.
\label{eq10}
\eeqa
The Eq.\eqref{eq10} appears as (detailed derivation is given in Appendix-\ref{apA}):
\beqa
\label{eq16}
S_{n} (\vec{\nabla}. \vec{v})+ \vec{\nabla}s_{+}. \vec{v}+ \vec{\nabla}. \Delta \vec{J}^{+}_{s}\ge0\,,
\eeqa
where
\beqa
S_{n}=s_{+}-\beta_{+}\,(\epsilon_{+}+P_{+})+\tilde{\alpha}_{+}\rho_{+}+\pi A_{\phi}\,.
\label{sn}
\eeqa

{\color{black}{The coefficient $S_{n}$ arises as the prefactor of $(\vec{\nabla}\cdot\vec{v})$ in the entropy production rate. In ordinary hydrodynamics (without $\phi$), thermodynamic consistency requires
\beqa
s - \beta(\epsilon+P) + \tilde{\alpha} \rho = 0\,.
\eeqa
In Hydro+, the additional mode $\phi$ modifies this balance: the entropy current divergence picks up an additional contribution from the last term, $\pi A_{\phi}$, originates from the coupling of $\phi$ to compression or expansion in Eq.~\eqref{slow}. It encodes how the slow mode changes the balance between entropy, enthalpy, and mass density compared to ordinary hydrodynamics. Physically, $S_{n}$ represents the net entropy contribution associated with fluid expansion or compression in the presence of the slow mode.}}

Since $S_n$ in Eq.\eqref{sn} is a zeroth-order thermodynamic quantity, whereas $\vec{\nabla}.\vec{v}$ is first order in the hydrodynamic gradient expansion, implies the term
$S_n(\vec{\nabla}.\vec{v})$ is linear in gradients. Moreover, the term $S_n(\vec{\nabla}\cdot\vec{v})$ changes sign between expansion and compression and hence cannot give a positive-definite entropy production for arbitrary flows. Therefore, to maintain thermodynamic consistency, one requires $S_n=0$.
This constraint relates the thermodynamic variables with the susceptibility $A_{\phi}$ as:
 \beqa
 \label{eq13}
 s_{+}=\beta_{+} (\epsilon_{+}+P_{+})-\tilde{\alpha}_{+} \rho_{+}-\pi A_{\phi}\,.
 \eeqa
 With $S_{n}=0$, the second law from Eq.\eqref{eqB1} takes the form (see Appendix-\ref{apB}):
\beqa
\label{eq21}
\pi F_{\phi}+ \vec{\nabla}. \Delta \vec{J}^{+}_{s}\ge0\,,
\eeqa

To guarantee the positivity condition of Eq.\eqref{eq21}, the second term will cancel with the first term. As the evolution equation of $\phi$ i.e., Eq.\eqref{slow} is already of relaxation type, the coupling of the first order of $\phi$ to the hydrodynamic field will be appropriate to describe the fluid. 

The choice of specific form of $\Delta \vec{J}^{s}_{+}$ must be in accordance with the principle of thermodynamics, maintaining the positive entropy production. The standard NS choice (without the $\phi$ parameter) provides the familiar heat flux term:
\beqa
\Delta \vec{J}^{s}_{+}=\beta_{+}{h_{+}\vec{v}}\,.
\eeqa
With the inclusion of $\phi$, the natural extension could be
\beqa
\Delta \vec{J}^{s}_{+}=\beta_{+}({h\vec{v}}+\phi\vec{v})+(\text{coupling terms})\,,
\eeqa
where the term \(\phi \vec{v}\) signifies the presence of \(\phi\) in the Gibbs-Duhem relation \(s_+ = \beta_+ (\epsilon_{+} + P_{+}) - \tilde{\alpha}_+ \rho_{+} - \pi \phi\). Any coupling term containing e.g., $\nabla\pi$ or $\nabla \phi$ may introduce additional gradient terms that would either break the positivity of entropy production or require additional transport coefficients beyond the scope of the first-order hydrodynamic theory. So, the minimal and sufficient choice for the coupling term is \(b \pi h \vec{v}\), which is first order in gradients, and is uniquely determined by the requirement that the divergence of \(\Delta \vec{J}^{s}_{+}\) cancel the gradient-dependent terms in \(F_\phi\), leaving only the positive-definite contribution \(\gamma \pi^2\). Therefore, the specific form of the entropy current correction is taken as
\beqa
\Delta \vec{J}^{s}_{+}=\beta_{+}({h\vec{v}}+\phi\vec{v}+b\pi h\vec{v})\,,
\eeqa
where \(b\) is a coupling strength between the slow mode and the heat flux. Its value may not fixed by thermodynamics alone but must be determined by matching to microscopic theory or experiment, ensuring that the entropy production remains non-negative.

The evolution equation of $\phi$ (see Appendix-\ref{apB} for detailed derivation) is:
\beqa
\frac{\partial \phi}{\partial t} + \vec{\nabla} \cdot (\phi \vec{v}) = -\gamma \pi + \frac{b}{T} \vec{\nabla} \cdot (h \vec{v}) - \frac{b h}{T^2} \vec{v} \cdot \vec{\nabla} T\,.
\label{NS4}
\eeqa 
This is the equation which governs the evolution of $\phi$.

To investigate density fluctuations, the hydrodynamic equations introduced above i.e., Eqs.~\eqref{NS1}, \eqref{NS2}, \eqref{NS3}, and \eqref{NS4} are linearized about the thermodynamic equilibrium state~\cite{Weinberg:1971mx}. This approach is valid when the system experiences only small deviations from equilibrium, allowing nonlinear terms in the fluctuations to be neglected. The resulting linearized hydrodynamic equations provide an effective framework for analyzing the fluctuation dynamics of the fluid.

Let $Q_0$ denote the equilibrium value of a thermodynamic variable $Q$. In the presence of weak perturbations, the variable may be expressed as
\begin{equation}
Q(\mathbf{r},t)=Q_0+\delta Q(\mathbf{r},t),
\end{equation}
where $\delta Q(\mathbf{r},t)$ represents a small fluctuation about the equilibrium value, 
satisfying $|\delta Q| \ll |Q_0|$. Here, $Q$ may correspond to any relevant thermodynamic quantity, 
such as the density, pressure, temperature, entropy, velocity field or $\phi$. 
Substituting these perturbed quantities into the hydrodynamic equations and retaining only terms linear in the fluctuations yields a closed set of linear equations. \textcolor{black}{For notational simplicity, we drop the `+' subscript and denote the thermodynamic variables without it throughout the following paper.}

For linearization about a homogeneous equilibrium state, each physical variable is written as its equilibrium value plus a small perturbation, e.g.,
\[
T=T_0+\delta T,\quad
\rho=\rho_0+\delta\rho,\quad
P=P_0+\delta P,\quad
h=h_0+\delta h,\quad
\phi=\phi_0+\delta\phi,\quad
\pi=\delta\pi,\quad
\vec v=\delta\vec v,
\]
with
\[
\vec v_0=\pi_{0}=0,\qquad
\nabla T_0=\nabla P_0=\nabla h_0=\nabla \rho_0=\nabla \phi_0=0.
\]
These conditions are imposed into the governing equations, and only terms linear in the perturbations are retained; products of perturbations, such as $\delta\vec v\cdot\nabla\delta T$, are neglected as second-order terms i.e., $\mathcal{O}(\delta^2)$. Consequently, the terms $\vec v\cdot\nabla T$, $\vec v\cdot\nabla P$, and $\vec v\cdot\nabla h$ vanish at linear order, while terms such as $\nabla^2\delta T$ are retained.

Following the linearization technique, the relevant four equations {\it{i.e.,}} Eq.~\eqref{NS1}, \eqref{NS2}, \eqref{NS3}, and \eqref{NS4} can be linearized as the following form:	
	\begin{subequations}
	\begin{eqnarray}
	\label{NSLa}
	0&=&\frac{\pd \delta \rho}{\pd t}+\rho_{0} \vec{\nabla}.\delta \vec{v}\,,  \\
	\label{NSLb}
	0&=& \rho_{0}\,\frac{\pd \delta v}{\pd t}+\vec{\nabla}\delta P-\Big(\zeta+\frac{4}{3}\eta\Big)\,\,\nabla^{2}\delta v\,,\\
	\label{NSLc}
	0&=& \rho_{0}C_{P}\frac{\pd \delta T}{\pd t}-\frac{\pd \delta P}{\pd t}-\chi \nabla^{2}\delta T\,,\\
	\label{NSLd}
 0&=&\frac{\partial\delta\phi}{\partial t}
+\Big(\phi_0-\frac{bh_{0}}{T_{0}}\Big)\vec\nabla\cdot\delta\vec v+\gamma \delta \pi\,,
	\eeqa
	\end{subequations}
where  in Eq.~\eqref{NSLc}, we have used $\delta \epsilon=C_{P}\,\delta T$, where $C_{P}$ is the specific heat at constant pressure. The $\rho_{0},\, T_{0},\,h_{0},\,\phi_{0}\,(=\bar{\phi})$ are the background number density, temperature, specific enthalpy, and relaxed slow mode respectively.

From Appendix-\ref{apC} (detailed derivation), the density fluctuation is obtained in the form:
\beqa
\label{eq23}
\delta {\rho}(\mathbf{k},\omega) 
&=& \left( -\mathbf{M}^{-1}_{11} + P_\rho \mathbf{M}^{-1}_{13} \right) \delta\rho(\mathbf{k}, 0) 
+ \rho_0 \mathbf{M}^{-1}_{12} \delta v(\mathbf{k}, 0) \nn\\
&&\quad + (P_T - \rho_0 C_P) \mathbf{M}^{-1}_{13} \delta T(\mathbf{k}, 0) 
+ \left( P_\phi \mathbf{M}^{-1}_{13} + \mathbf{M}^{-1}_{14} \right) \delta\phi(\mathbf{k}, 0)\,,
\eeqa
where
\beqa
P_{\rho}=\big(\frac{\partial P}{\partial \rho}\big)_{T,\,\phi},\,\,\,P_{T}=\big(\frac{\partial P}{\partial T}\big)_{\rho,\,\phi},\,\,\,P_{\phi}=\big(\frac{\partial P}{\partial \phi}\big)_{\rho,\,T}\,.
\eeqa
Consequently, the density correlator (see Appendix- \ref{apC}) is evaluated as:
\beqa
\mathcal{S}_{\rho\rho}(\mathbf{k}, \omega) = \frac{\mathcal{N}_3 \omega^3 + \mathcal{N}_2 \omega^2 + \mathcal{N}_1 \omega + \mathcal{N}_0}{\mathcal{D}_4 \omega^4 + \mathcal{D}_3 \omega^3 + \mathcal{D}_2 \omega^2 + \mathcal{D}_1 \omega} \,.
\eeqa
The aim of the present work is to study the the effects of $\phi$ on $\mathcal{S}_{\rho\rho}$.
\section{Results and discussion}
\label{sec3}
The dynamic structure factor, \(\mathcal{S}_{\rho\rho}(\bm{k}, \omega)\), is one of the key measurable observables for probing the critical dynamics of the fluid. Within the Hydro+ framework, we numerically evaluate \(\mathcal{S}_{\rho\rho}(\bm{k}, \omega)\) to elucidate the distinctive signatures of the out-of-equilibrium mode (OEM), \(\phi\). The equation of state for He\(^{4}\) is adapted from Ref.~\cite{Bezverkhy} and used the scaling laws
from Ref.\cite{Kadanoff} 
for the variation of transport coefficeints ($\eta$, $\zeta$ and $\chi$) and  response
functions ($C_p$ and $(C_v$) with temperature near the critical point to generate the results displayed in
this work. 

\begin{figure}
	\centering
	\includegraphics[width=0.31 \textwidth]{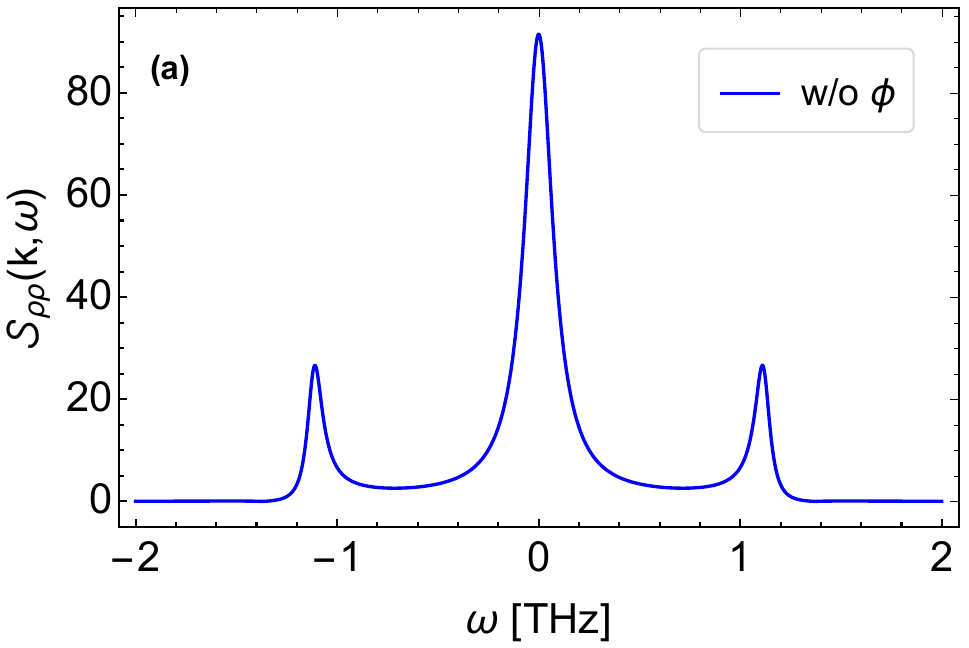}
	\includegraphics[width=0.32 \textwidth]{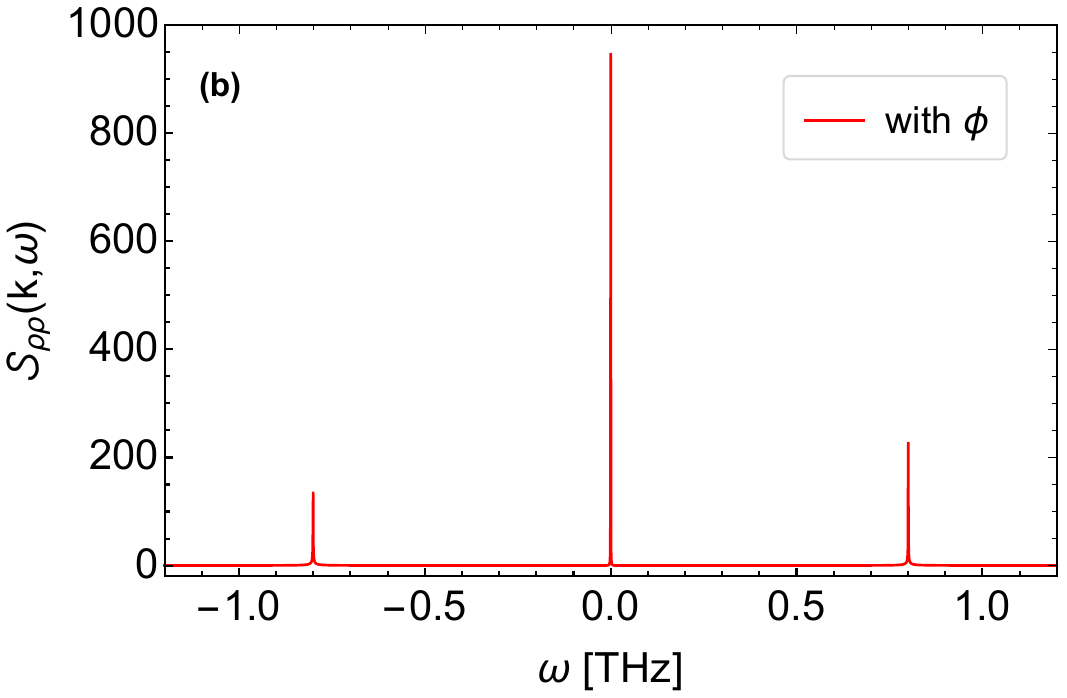}
	\includegraphics[width=0.31 \textwidth]{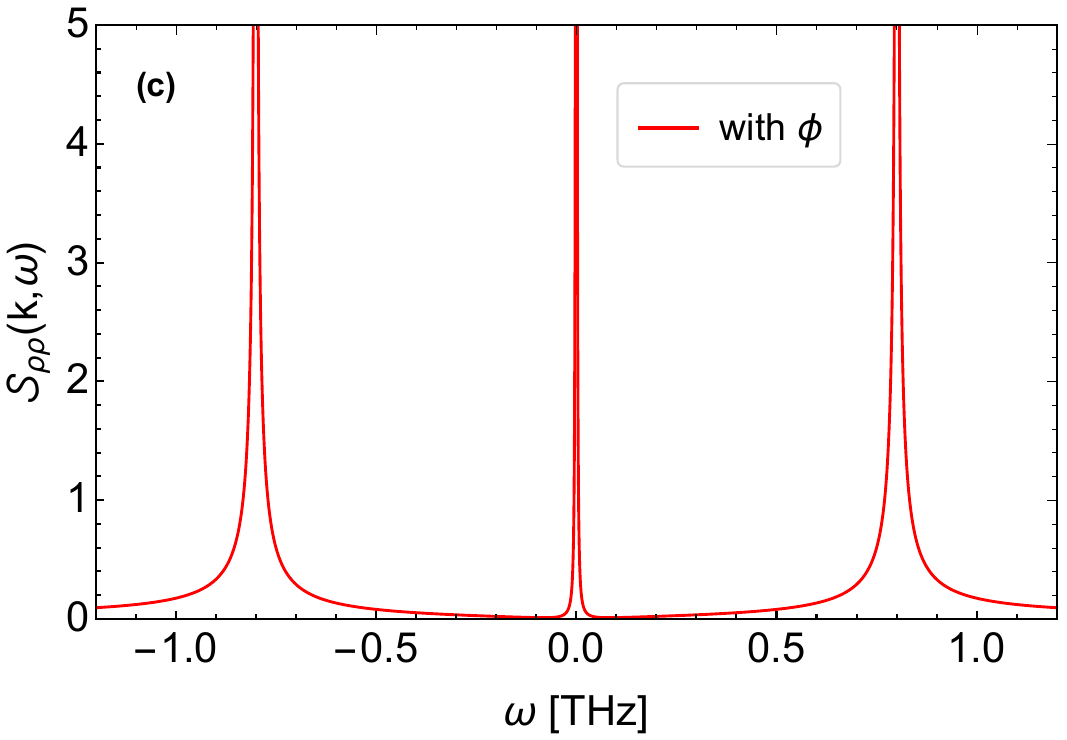}
\caption{{The variation of the $\Snn$ with $\omega$ is shown when the system is away from the critical point.  
Here $\Snn$ is plotted at $T=20$ K by choosing $k=1\,\AA^{-1}$, and 
(a) $\bar{\phi}=0,\,b=0$, (b) $\bar{\phi}=0.01\,h_{0}, \,b=0.001 \,K$, 
and (c) $\bar{\phi}=0.01\,h_{0},\,b=0.001\,K$. The panel (c) is 
same as panel (b) with limited ordinate values.}}
	\label{fig1}
\end{figure}

To establish a baseline for comparison, we first examine the behavior of the dynamic structure factor $\mathcal{S}_{\rho\rho}(\bm{k}, \omega)$ in a regime away from the critical point. Fig.~\ref{fig1} displays $\mathcal{S}_{\rho\rho}(\bm{k}, \omega)$ at $T = 20$ K, which is substantially above the critical temperature of $T_c = 5.2$ K for the He$^4$ fluid. Fig.\ref{fig1} presents the characteristic structure of $\mathcal{S}_{\rho\rho}(\bm{k}, \omega)$ is plotted as a function of frequency $\omega$ for $k =1$ $\AA^{-1}$. Fig.\ref{fig1}(a) shows the $\Snn$ without the incorporation of the OEM. Fig.\ref{fig1}(b) shows 
the variation of $\mathcal{S}_{\rho\rho}$ with $\omega$ with the inclusion of $\phi$.
while Fig.\ref{fig1}(c) panel employs results with lower range of ordinate values
to make the detailed lineshapes of the individual peaks of Fig.\ref{fig1}(b) visible.

Under these non-critical conditions, $\mathcal{S}_{\rho\rho}(\bm{k}, \omega)$ exhibits the conventional three-peak structure: a central Rayleigh peak arising from thermal diffusion \cite{Rayleigh1881}, flanked symmetrically by two Brillouin peaks corresponding to propagating 
sound modes \cite{FlerryandBoon1969}. The peaks maintain well-defined Lorentzian profiles with appreciable  widths, indicating rapid dissipations. With the inclusion of the OEM ($\bar{\phi}=0.01\,h_{0}$, $b=0.001\,K$, where $h_{0}\approx 10^{5}$ Jkg$^{-1}$), the structural pattern of the spectrum remains unaltered, but the position of the Brillouin peaks have shifted towards $\omega=0$. This shift is attributed to the enhanced non-equilibrium fluctuations introduced by the slow mode. As the fluctuation amplitude increases, the pressure response to density perturbations becomes slower, reducing the effective elastic restoring force associated with acoustic propagation. Consequently, the effective sound speed decreases, causing the Brillouin peaks to move closer to the central Rayleigh peak.
Introduction of $\phi$ reduces the width of the distribution indicating slower decay [Fig. \ref{fig1}(b) and \ref{fig1}(c)]. 
The reduction of widths indicate the deceleration of dissipation.

\begin{figure}
	\centering
	\includegraphics[width=0.45 \textwidth]{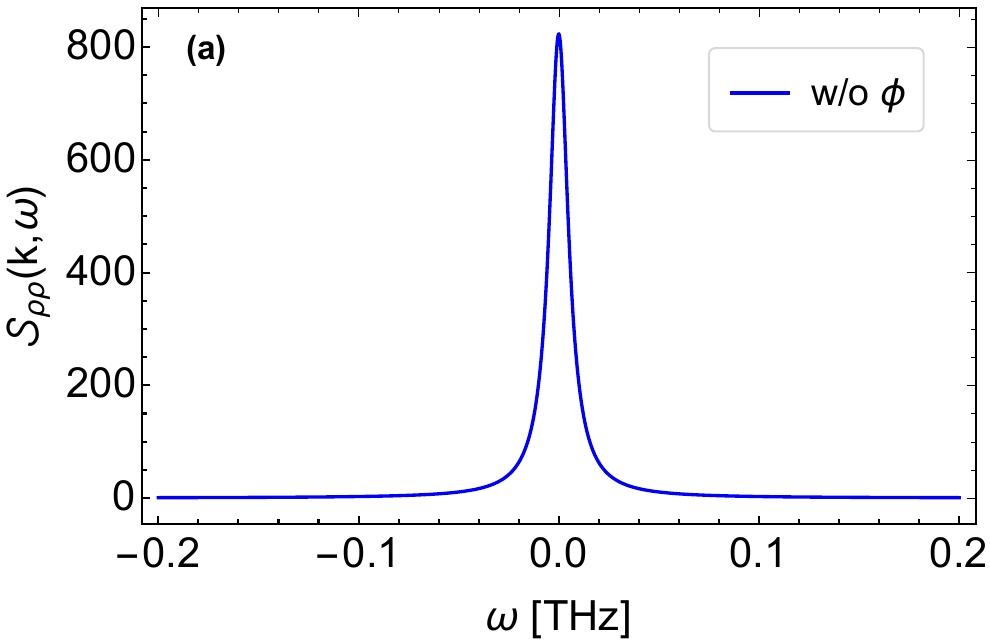}
	\includegraphics[width=0.467 \textwidth]{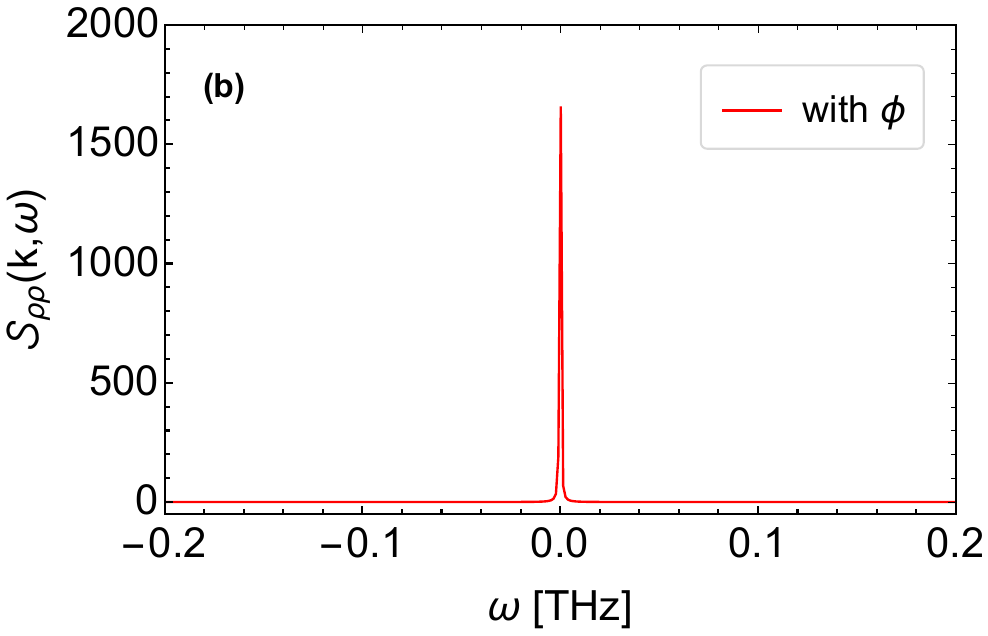}
\caption{The variation of the $\Snn$ with $\omega$ is shown when the system is near the critical point.  
Here $\Snn$ is plotted at $T=6$ K by choosing $k=1\,\AA^{-1}$, $\bar{\phi}=0.01\,h_{0}$, and $b=0.001\,K$. 
The results in panel (a) stands for $\phi=0$
and panel (b) shows results for $\phi \neq 0$. 
The scaling laws of the transport coefficients~\cite{Kadanoff} are presented in Appendix-\ref{apD}.}
	\label{fig2}
\end{figure} 


\begin{figure}[htbp]
	\centering
	\includegraphics[width=0.55 \textwidth]{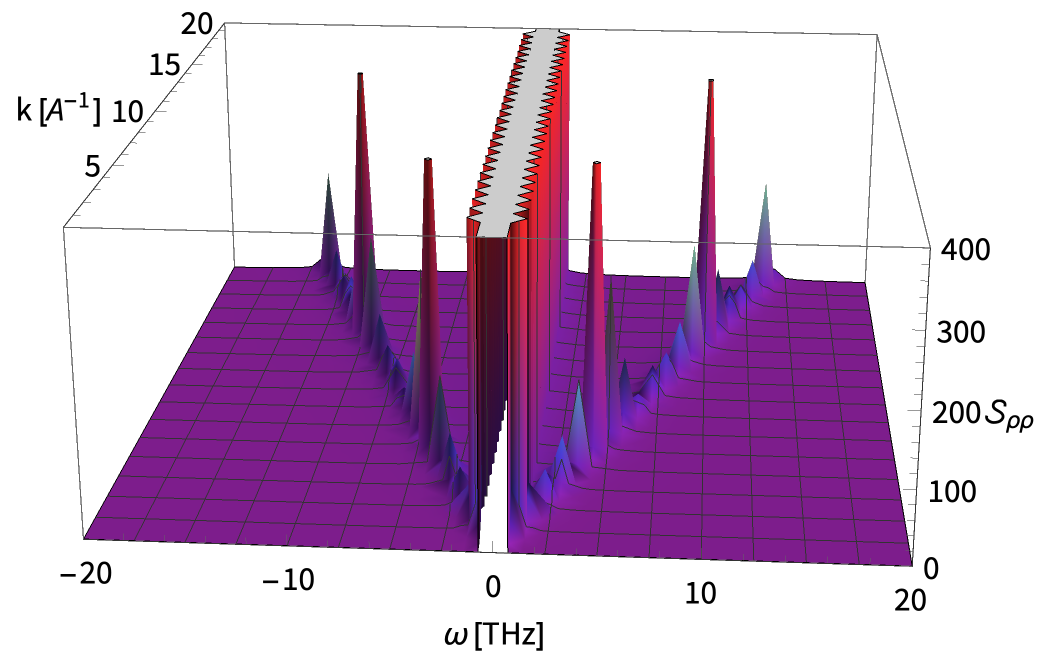}
	\caption{A 3D plot of $\mathcal{S}_{\rho\rho}(\bm{k}, \omega)$ with $\omega$ and $k$ is shown at $T=20$ K by choosing $\bar{\phi}=0.01\,h_{0}$, and $b=0.001$. We have scaled down the plot to clearly visualize the Brillouin peaks.}
	\label{fig:3d_omega_k}
\end{figure}

Fig.~\ref{fig2} presents the characteristic line-shape of $\mathcal{S}_{\rho\rho}(\bm{k}, \omega)$ as 
a function of frequency $\omega$ at a
fixed wave number $k = 1$ \AA$^{-1}$ and a temperature of $T = 6$ K, which is closer to the critical 
temperature of $T_c = 5.2$ K. 
In both Figs. \ref{fig2}(a) and (b) the pattern remains same, and we see only the Rayleigh peaks as expected. Near the critical point, 
the density fluctuation becomes very large, indicating a significant increase in the compressibility 
$\kappa=-\frac{1}{V}(\frac{\pd V}{\pd P})_T$. Consequently, $c_{s}^{2}=\frac{1}{\rho\,\kappa}$ vanishes, which results in vanishing of the 
Brillouin peaks. Thus vanishing of Brillouin peaks in the structure factor is one of the feature 
of the critical point indicating strong attenuation of 
sound mode~\cite{Stanley}. \textcolor{black}{The significant reduction in the width of the Rayleigh 
peak in presence of $\phi$ indicates slower damping  of the fluctuations, also known as critical slowing down.}

To further investigate the dispersion and spectral weight distribution, we plot $\mathcal{S}_{\rho\rho}$ in three dimensions as a function of both $\omega$ and $k$ at $T=20$ K, as shown in Fig.~\ref{fig:3d_omega_k}. The 3D visualization allows us to track the evolution of the Brillouin peaks with the wave number. At higher wave numbers ($k$), the Brillouin peaks are observed to move farther from the Rayleigh peak, consistent with the dispersion relation  $\omega=c_{s} k$ governing sound modes in the medium. 

\begin{figure}[htbp]
	\centering
	\includegraphics[width=0.5 \textwidth]{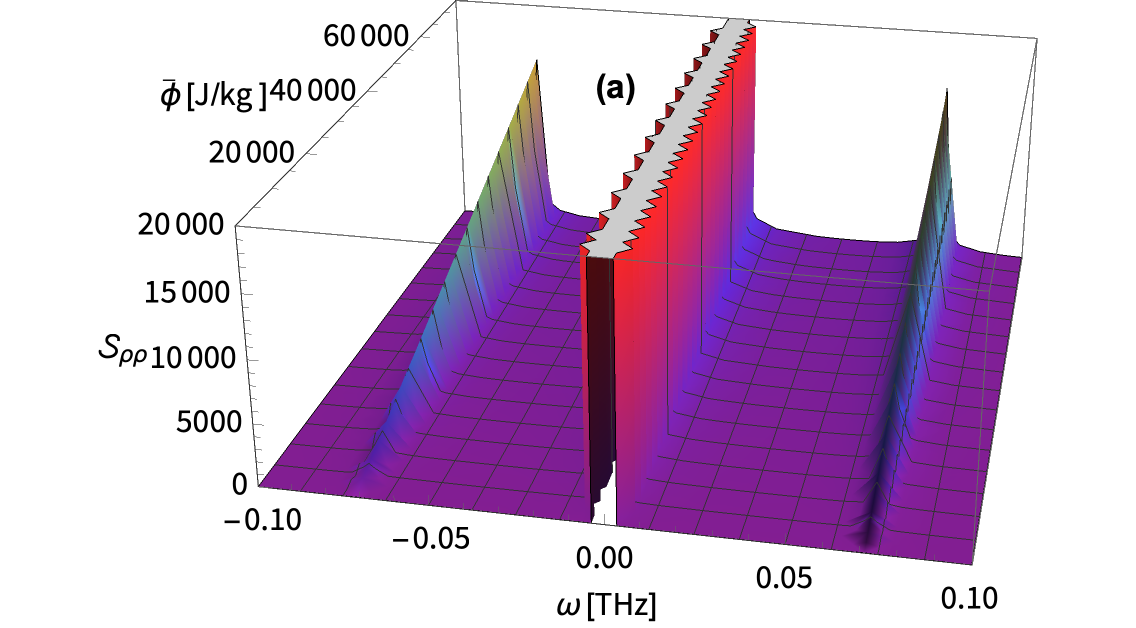}
	\includegraphics[width=0.45 \textwidth]{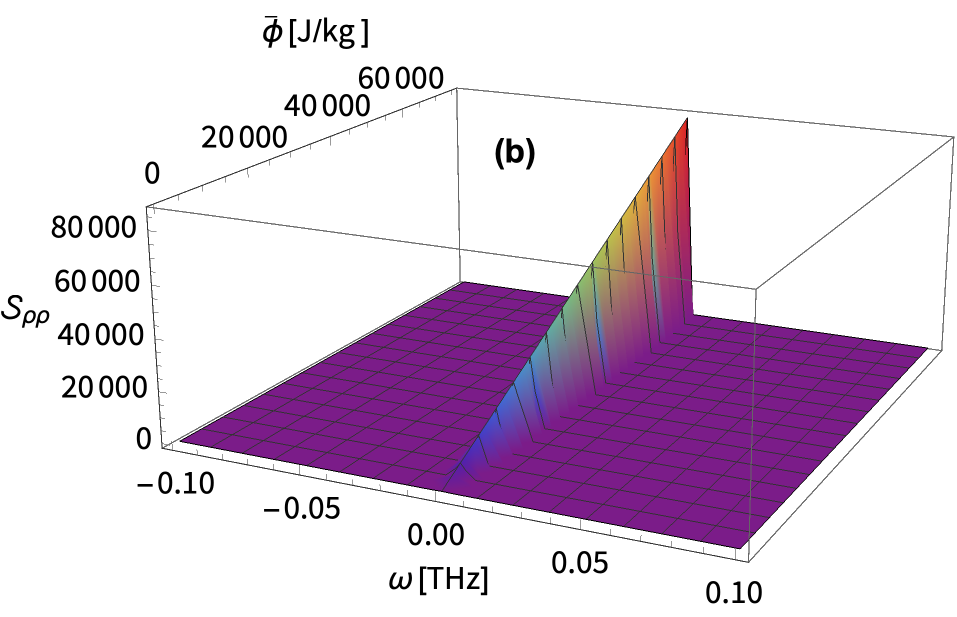}
	\caption{A 3D plot of $\mathcal{S}_{\rho\rho}(\bm{k}, \omega)$ with $\omega$ and $\phi$ is shown at $T=20$ K. 
Left panel displays the results with a cut-off in  $\mathcal{S}_{\rho\rho}$ to make the Brillouin peaks visible. 
The right panel shows the same results  without any cut-off.}
	\label{fig4}
\end{figure}

The role of the slow mode $\phi$ is explored in Fig.~\ref{fig4}, which depicts the variation of $\mathcal{S}_{\rho\rho}$ with $\omega$ and the equilibrium 
value of the slow mode, $\bar{\phi}$. In the left panel the ordinate is scaled to make the Brillouin peaks visible, and the right panel, showing the full spectrum, demonstrate that the intensity and character of the peaks are highly sensitive to the value of $\bar{\phi}$. With increasing $\phi$ ($1\%\,\le \bar{\phi}/h_{0}\le 75\%$), the heights of the Brillouin peaks as well as the Rayleigh peak increases. The peak height grows monotonically while its position remains unchanged. The change in the peak heights is a consequence of the redistribution of spectral weight due to the coupling between the hydrodynamic variables and the slow non-equilibrium mode. The slow mode does not create new hydrodynamic modes; rather, it modifies how the fluctuation energy is distributed among the existing modes.

When $\phi$ is introduced into the system, it acts as an additional thermodynamic degree of freedom capable of storing a portion of the fluctuation energy. As a consequence, the energy associated with a density fluctuation is no longer partitioned solely between the thermal diffusive mode and the propagating sound modes. Instead, a part of this energy is temporarily trapped in the slow mode before being released through relaxation. 
As the magnitude of $\phi$ mode increases, the slow mode stores more fluctuation energy. 
Because the slow mode relaxes diffusively, increasing $\phi$ channels more fluctuation energy into the low-frequency regime. Consequently, the central Rayleigh peak becomes relatively more prominent than the Brillouin peaks. As a result, the lifetime of thermal fluctuation significantly increases, causing the critical slowing down. Thus, one can argue that the $\phi$ mode acts as a source of fluctuation, and redistribute it in the power spectrum to enhance the magnitudes of the peaks.

\begin{figure}[htbp]
	\centering
	\includegraphics[width=0.45 \textwidth]{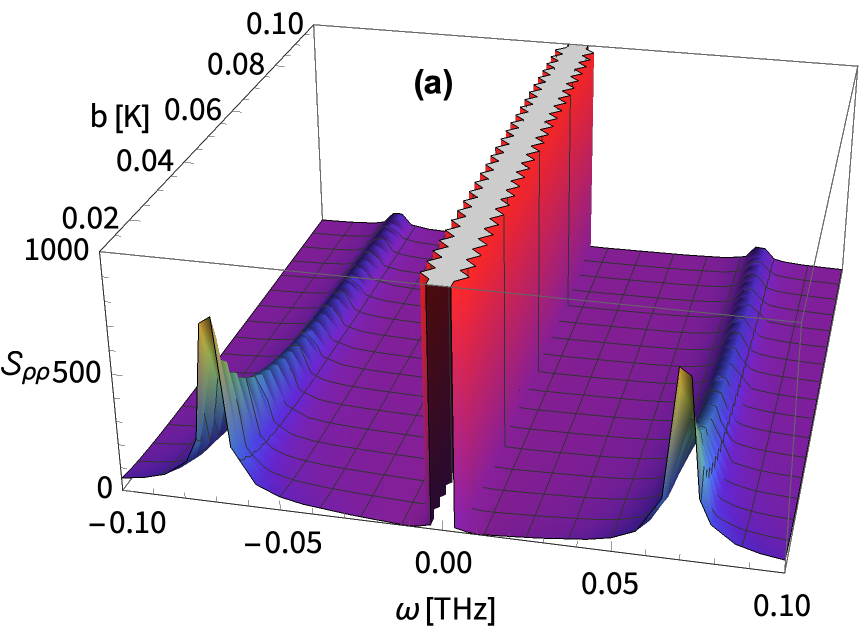}
	\includegraphics[width=0.5 \textwidth]{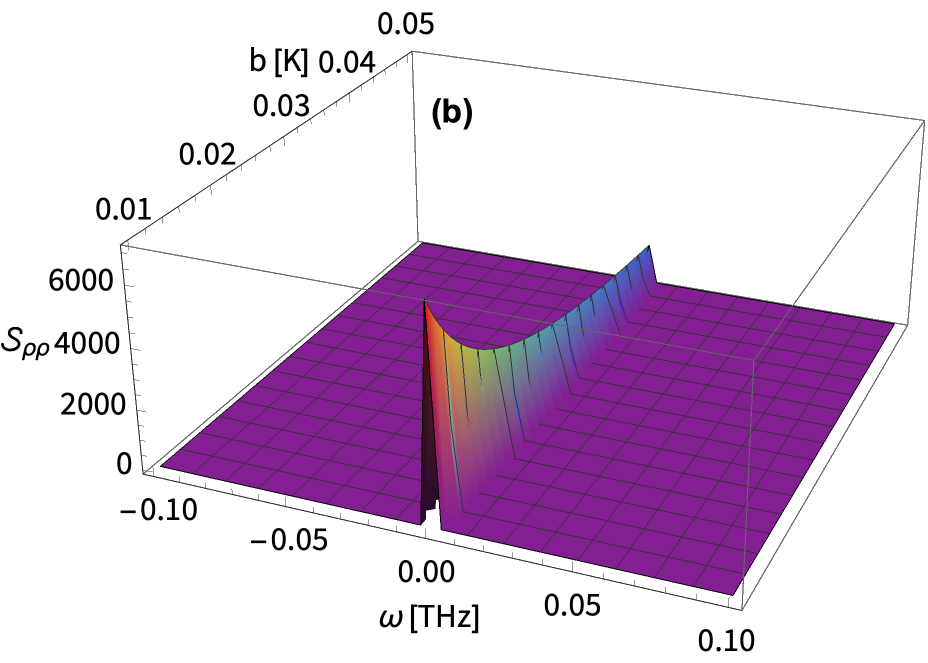}
	\caption{A 3D plot of $\mathcal{S}_{\rho\rho}(\bm{k}, \omega)$ with $\omega$ and $b$ is shown at $T=20$ K. 
The ordinate in the left panel is scaled to clearly visualize the Brillouin peaks. The right panel is the same plot but with whole range of coordinates.}
	\label{fig5}
\end{figure}

To examine the influence of the coupling between the slow non-equilibrium mode and the heat flux, 
we investigate the effects of the coupling between $\phi$ and heat flux
on the  Brillouin and Rayleigh components of the dynamic structure factor.
The parameter is varied over the range $0.001\,K\le b \le 0.10\,K$, while the keeping $\bar{\phi}=0.01\,h_{0}$.

Fig.\ref{fig5}(a) illustrates the dependence of the Brillouin component of the dynamic structure factor on the frequency $\omega$ and the coupling parameter b. The spectrum exhibits two symmetric Brillouin peaks corresponding to the propagating sound modes located at finite frequencies. As the coupling parameter increases, the intensity of the Brillouin peaks decreases monotonically, whereas their positions remain unchanged. This indicates that the coupling between the slow mode and the heat flux primarily affects the amplitude of the acoustic fluctuations without significantly modifying the equilibrium sound velocity, $c_{s}=\omega/k$.

The dependence of Rayleigh peak on $b$ is presented in Fig. \ref{fig5}(b). The central Rayleigh peak also exhibits a 
similar pattern as that of the Brillouin peaks. The peak remains centered at $\omega=0$, confirming that the 
associated fluctuations remain purely diffusive in nature. The coupling parameter $b$ controls the efficiency 
with which the $\phi$ mode exchanges energy with the hydrodynamic modes. As $b$ increases, the coupling becomes 
stronger, enabling the slow mode to dissipate the fluctuation more efficiently. As a result, the lifetime of 
both thermal and acoustic fluctuations is reduced, leading to lower peak heights in the dynamic structure factor.
\begin{figure}[htbp]
	\centering
	\includegraphics[width=0.47 \textwidth]{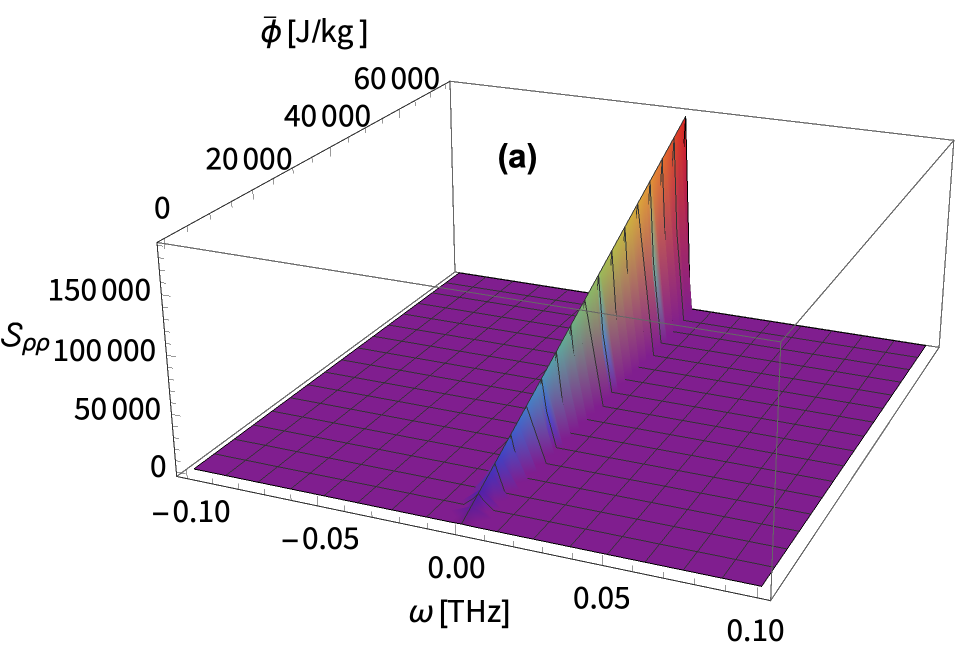}
	\includegraphics[width=0.47 \textwidth]{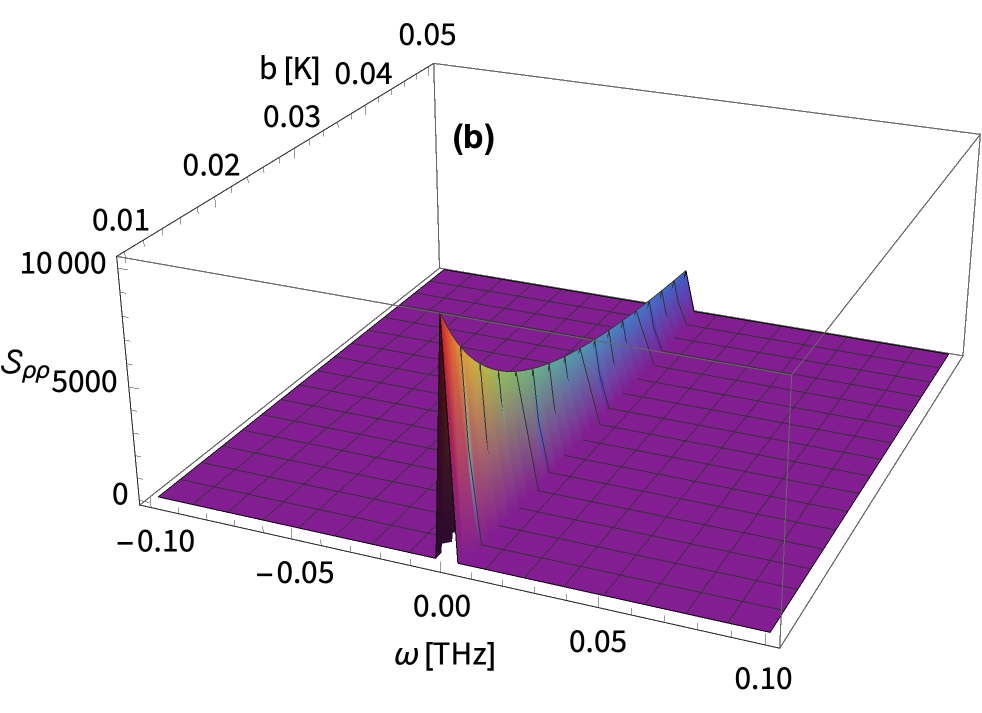}
	\caption{Left panel shows a 3D plot of $\mathcal{S}_{\rho\rho}(\bm{k}, \omega)$ with $\omega$ and $\phi$ is shown at $T=6$ K (near the CP). Right panel shows the 3D plot of $\mathcal{S}_{\rho\rho}(\bm{k}, \omega)$ with $\omega$ and $b$ is shown at $T=6$ K.}
	\label{fig6}
\end{figure}

Fig.\ref{fig6} shows the plots of $\Snn$ near the critical region. The Brillouin peaks disappear, and we observe the 
effects of $\phi$ and $b$ on the Rayleigh peak only. 
Here the $\bar{\phi}$ is varied in the range $0.01\,h_{0}\le \bar{\phi}\le 0.75\,h_{0}$, where 
$h_{0}\approx10^{5} $ Jkg$^{-1}$, 
is the value of the equilibrium enthalpy density near the critical temperature $T=6$ K.
With increasing $\phi$, the the Rayleigh peak increases monotonically, whereas with $b$, the magnitude of the peak decreases. The behaviour of the Rayleigh peak with variation of $b$ is similar to the situation when the system was well away from the critical region.
\begin{figure}[htbp]
	\centering
	\includegraphics[width=0.65 \textwidth]{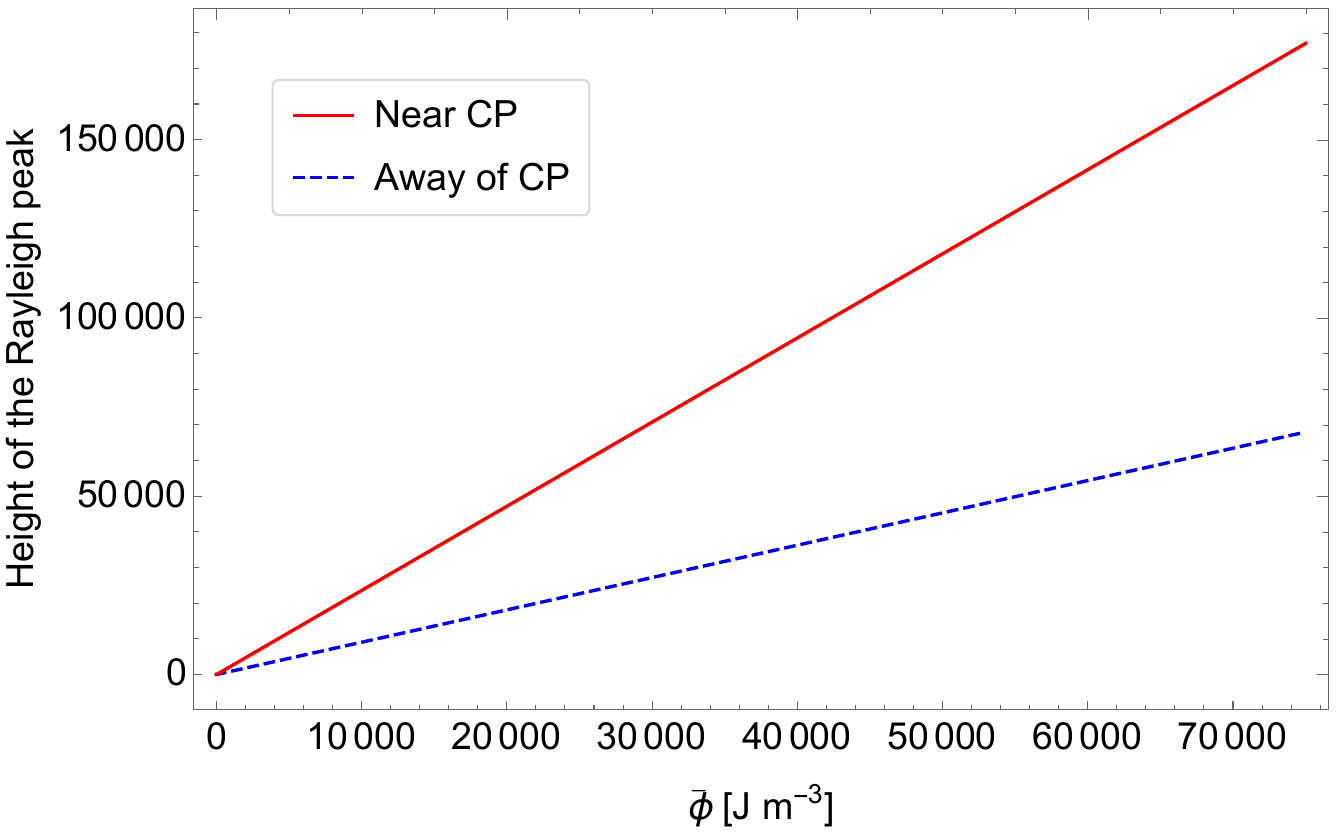}
	\caption{Figure shows the change in the height of the Rayleigh peak with the variation of $\phi$. This is the 2D plot extracted from Fig.\ref{fig4}(b) and Fig.\ref{fig6}(a).}
	\label{fig7}
\end{figure}
Fig.\ref{fig7} shows the variation of the Rayleigh peak height with the $\phi$ mode for the fluid near and away from the critical point. 
It is seen that the Rayleigh peak height increases as $\phi$ increases in both cases. 
However, the increase is much larger near the critical point than when the system is away from it. 
This is because the slow mode becomes more pronounced near the critical point due to critical slowing down. As a result, more fluctuation is stored in the slow mode, which enhances the thermal fluctuations and increases the height of the Rayleigh peak. Away from the critical point, the slow mode relaxes more quickly, so its effect on the thermal fluctuations is weaker, leading to a smaller increase in the Rayleigh peak height. These results show that the slow non-equilibrium mode has a much stronger influence on the thermal fluctuations near the critical point than in the non-critical region.

\section{Summary, conclusion and outlook}
\label{sec4}
In this work, we have extended the Navier-Stokes hydrodynamic framework by incorporating an out-of-equilibrium mode (OEM) to describe a fluid near the critical point. Starting from a generalized entropy current, we derived the corresponding constitutive relations while ensuring consistency with the second law of thermodynamics. The linearized hydrodynamic equations were then used to obtain the dynamic structure factor to understand the criticality in a fluid.

Our analysis demonstrate that the OEM produces noticeable modifications in the dynamical response of the fluid. The introduction of the slow-mode parameter $\phi$ enhances the thermal fluctuations, resulting in a significant increase in the height of the Rayleigh peak. At the same time, the Brillouin peaks shift towards lower frequencies, indicating a reduction in the effective speed of sound due to the slower response of pressure fluctuations. The increase in the Rayleigh peak is much more pronounced near the critical point than away from it, highlighting the important role of critical slowing down in amplifying thermal fluctuations. Furthermore, increasing the coupling parameter $b$, which characterizes the interaction between the slow mode and the heat flux, suppresses the heights of both the Rayleigh and Brillouin peaks. This behaviour suggests that a stronger coupling introduces an additional relaxation channel, leading to enhanced damping of both thermal and acoustic fluctuations. Overall, the OEM redistributes the fluctuation energy among the hydrodynamic modes and significantly modifies the dynamic structure factor.

There are several possible directions for future work. The present analysis can be extended by considering nonlinear fluctuations. It would also be interesting to study the role of multiple slow modes, investigate the dependence of transport coefficients on the slow mode, and compare the theoretical predictions with experimental measurements of light or neutron scattering in critical fluids such as liquid 
He$^{4}$. Such studies may provide a deeper understanding of non-equilibrium critical dynamics and the role of slow modes in condensed matter systems.

\section{Appendix}
\appendix
\section{Evaluation of $S_{n}$}
\label{apA}
In this appendix we derive the following expression
\[
S_{n}=s_{+}-\beta_{+}(\epsilon+P)+\tilde{\alpha}_{+}\rho+\pi A_{\phi}.
\]
To begin with:
  \begin{equation}
  ds_{+}=\beta_{+}\,d\epsilon-\tilde{\alpha}_{+}\,d\rho-\pi\,d\phi .
  \label{eq:entropy_diff}
  \end{equation}
Using the material derivate:
  \beqa
   \frac{D}{Dt}
  &=&\frac{\partial }{\partial t}+\vec v\!\cdot\nabla\,,
  \eeqa
such that Eqs.\eqref{NS1}, \eqref{NS3}, and \eqref{slow} appear as:
  \begin{eqnarray}
  \label{A3}
  \frac{D\rho}{Dt}&=&-\rho\,\nabla\!\cdot\vec v, \\
  \label{A4}
  \frac{D\epsilon}{Dt}&=&-(\epsilon+P)\,\nabla\!\cdot\vec v
   \;+\;\mathcal O(\nabla^2),\\
   \label{A5}
  \frac{D\phi}{Dt}  &=&-F_{\phi}-A_{\phi}(\nabla\!\cdot\vec v).
  \end{eqnarray}

From \eqref{eq:entropy_diff} we have
\begin{equation}
\frac{Ds_{+}}{Dt}
=\beta_{+}\frac{D\epsilon}{Dt}
 -\tilde{\alpha}_{+}\frac{D\rho}{Dt}
 -\pi \frac{D\phi}{Dt}.
 \label{A6}
\end{equation}
Inserting the equations \eqref{A3}, \eqref{A4}, and \eqref{A5} in Eq.\eqref{A6}, we have
\begin{equation}
\frac{Ds_{+}}{Dt}
=\big[-\beta_{+}(\epsilon+P)+\tilde{\alpha}_{+}\rho+\pi A_{\phi}\big]
 (\nabla\!\cdot\vec v)
 + \pi F_{\phi}
 + \mathcal O(\nabla^2).
\label{eq:Ds}
\end{equation}

The entropy production equation gives:
\beqa
\pd_{t}s_{+}+\vec{\nabla}.\vec{J}^{s}_{+}&\ge& 0,\nn\\
 \frac{Ds_{+}}{Dt}+s_{+}(\vec{\nabla}.\vec{v})+\vec{\nabla}.\Delta\vec{J}^{s}_{+}&\ge&0\,.
 \label{eqAA}
\eeqa

Substituting Eq.\eqref{eq:Ds} into \eqref{eqAA}, we get
\beqa
\big[s_{+}-\beta_{+}(\epsilon+P)+\tilde{\alpha}_{+}\rho+\pi A_{\phi}\big]
 (\nabla\!\cdot\vec v)
 + \pi F_{\phi}+\vec{\nabla}.\Delta\vec{J}^{s}_{+}+\mathcal O(\nabla^2)\ge0.
\eeqa
Neglecting $\mathcal O(\nabla^2)$, we write
\beqa
S_{n}(\nabla\!\cdot\vec v)
+\pi F_{\phi}
+ \vec{\nabla}.\Delta\vec{J}^{s}_{+}\ge0,
\eeqa
we identify
\begin{equation}
{S_{n}=s_{+}-\beta_{+}(\epsilon+P)+\tilde{\alpha}_{+}\rho+\pi A_{\phi}}.
\end{equation}
In complete equilibrium ($\pi=0$, $A_{\phi}=0$), and we obtain the usual thermodynamic relation:
\beqa
s=\beta(\epsilon+P)-\tilde{\alpha} \rho\,,
\eeqa
which ensures $S_{n}=0$, as required.
Near a critical point, however, the additional term $\pi A_{\phi}$ modifies this balance, encoding the coupling of the slow mode to fluid expansion.

\section{Evaluation of $F_{\phi}$}
\label{apB}
 As discussed in Sec.\ref{sec2}, with $S_{n}=0$, the entropy production equation with the remaining terms is:
\beqa
\label{eqB1}
\pi F_{\phi}+ \vec{\nabla}. \Delta \vec{J}^{+}_{s}\ge0\,,
\label{eqB1}
\eeqa
with
\beqa
\label{eqB3}
\Delta \vec{J}^{s}_{+}=\beta_{+}({h\vec{v}}+\phi\vec{v}+b\pi h\vec{v})\,,
\eeqa
Evaluating:
\beqa
\label{eqB3}
\vec{\nabla}\cdot\Delta\vec{J}^{+}_{s}
&=&
(\vec{\nabla}\beta_{+})\cdot(h\vec{v})
+(\vec{\nabla}\beta_{+})\cdot(\phi\vec{v})
+b\pi(\vec{\nabla}\beta_{+})\cdot(h\vec{v})
\nonumber\\
&+&\beta_{+}\vec{v}\cdot\vec{\nabla}h
+\beta_{+}\vec{v}\cdot\vec{\nabla}\phi
+b\beta_{+}\vec{v}\cdot
\left(\pi\vec{\nabla}h+h\vec{\nabla}\pi\right)
\nonumber\\
&+&\beta_{+}(h+\phi+b\pi h)
(\vec{\nabla}\cdot\vec{v}).
\eeqa
Using Eq.\eqref{eqB3} in Eq.\eqref{eqB1}, we obtain
\beqa
&& \pi
\Big[
F_\phi
+b(\vec{\nabla}\beta_{+})\cdot(h\vec{v})
+b\beta_{+}\vec{v}\cdot\vec{\nabla}h
+b\beta_{+}h(\vec{\nabla}\cdot\vec{v})
\Big]+(\vec{\nabla}\beta_{+})\cdot(h\vec{v})
+(\vec{\nabla}\beta_{+})\cdot(\phi\vec{v})\nn\\
&&+\beta_{+}\vec{v}\cdot\vec{\nabla}h
+\beta_{+}\vec{v}\cdot\vec{\nabla}\phi
+\beta_{+}(h+\phi)(\vec{\nabla}\cdot\vec{v})+b\beta_{+}h\vec{v}.\vec{\nabla}\pi \ge 0.
\eeqa
For the above equation to be positive definite, we require:
\beqa
F_\phi
+b(\vec{\nabla}\beta_{+})\cdot(h\vec{v})
+b\beta_{+}\vec{v}\cdot\vec{\nabla}h
+b\beta_{+}h(\vec{\nabla}\cdot\vec{v})=\gamma \pi\,.
\eeqa
where $\gamma>0$ is some constant.

We finally get the form of $F_{\phi}$ as:
\begin{equation}
F_\phi
=
\gamma\pi
-b\left[
(\vec{\nabla}\beta_{+})\cdot(h\vec{v})
+\beta_{+}\vec{v}\cdot\vec{\nabla}h
+\beta_{+}h(\vec{\nabla}\cdot\vec{v})
\right],
\label{fphi}
\end{equation}

Now, by comparing Eq.\eqref{eq8} and \eqref{eq13}, we have $A_{\phi}=\phi$. With the substitution of $F_{\phi}$ from above equation and $A_{\phi}=\phi$ in the Eq.\eqref{slow}, we obtain the complete equation for $\phi$ as follows:
\beqa
\label{slowfinal1}
\frac{\pd \phi}{\pd t}+\vec{\nabla}.(\phi\vec{v})&=&-\gamma\pi
+b\left[
(\vec{\nabla}\beta_{+})\cdot(h\vec{v})
+\beta_{+}\vec{v}\cdot\vec{\nabla}h
+\beta_{+}h(\vec{\nabla}\cdot\vec{v})
\right]\,.
\eeqa
Using Eq.\eqref{NS3} and using $\beta=1/T$ into the above equation, we obtain
\beqa
\frac{\partial \phi}{\partial t} + \nabla \cdot (\phi \vec{v}) = -\gamma \pi + \frac{b}{T} \nabla \cdot (h \vec{v}) - \frac{b h}{T^2} \vec{v} \cdot \nabla T
\label{B8}
\eeqa
This is the equation which governs the evolution of $\phi$.

\section{Evaluation of $\delta \rho(\vec{k},\,\omega)$ and the $\mathcal{S}_{\rho \rho}(\vec{k},\omega)$}
\label{apC}
By applying Fourier-Laplace transformation {\textit{i.e.}}, $\displaystyle{\lim_{x \to \infty}} \int d^3x\int_{0}^{\infty}dt\,\exp[i({\omega t-\vec{k}\cdot\vec{x}})]$ from the left in the set of equations Eqs.\eqref{NSLa}-\eqref{NSLd} and performing the integration, we get the linearized set of Fourier-Laplace transformed equations in the $\omega-k$ space:	
\begin{subequations}
\label{FLsystem}
\begin{eqnarray}
\label{c1}
&& i\omega \, \delta{\rho}(\vec{k},\omega) - i k \rho_0 \, \delta{v}(\vec{k},\omega) = -\delta\rho(\mathbf{k}, 0) \\
&&  i k P_\rho \, \delta{\rho}(\vec{k},\omega) + [k^2 \nu -i\omega \rho_0] \delta{v}(\vec{k},\omega) + i k P_T \, \delta{T}(\vec{k},\omega) + i k P_\phi \, \delta{\phi}(\vec{k},\omega) = \rho_0 \, \delta v(\mathbf{k}, 0)\,, \\
\label{c2}
&& -i\omega P_\rho \, \delta{\rho}(\vec{k},\omega) + \left[ i\omega \rho_0 C_P - i\omega P_T - \chi k^2 \right] \delta(\vec{k},\omega){T} - i\omega P_\phi \, \delta{\phi}(\vec{k},\omega)\,, \nn\\ 
\label{c3}
&&=P_\rho \, \delta\rho(\mathbf{k}, 0) + \left[ P_T - \rho_0 C_P \right] \delta T(\mathbf{k}, 0) + P_\phi \, \delta\phi(\mathbf{k}, 0)\,, \\
\label{c4}
&&\gamma P_\rho \, \delta{\rho} (\vec{k},\,\omega)+ i k \xi \delta{v}(\vec{k},\,\omega)+ \gamma P_T \, \delta{T}(\vec{k},\,\omega) + \left( -i\omega + \gamma P_\phi \right) \delta{\phi}(\vec{k},\,\omega) = \delta\phi(\mathbf{k}, 0)\,,
\end{eqnarray}
\end{subequations}
where 
\beqa
\nu=\zeta+\frac{4}{3}\eta,\,\,\,\,\,\,\,\xi= \phi_0 - \frac{b h_0}{T_0} \,.
\eeqa

In Eqs.~\eqref{c1}-\eqref{c4}, we have used
\beqa
 \delta P=\big(\frac{\partial P}{\partial \rho}\big)\,\delta \rho+\big(\frac{\partial P}{\partial T}\big)\,\delta T+\big(\frac{\partial P}{\partial \phi}\big)\,\delta \phi\,=P_{\rho}\delta \rho+P_{T}\delta T+P_{\phi}\delta \phi\,.
\eeqa

The Eqs.~\eqref{c1}-\eqref{c4} can be expressed in a matrix form as:
\beqa
\label{c4}
\mathds{M} \,\,\delta\mathcal{Z}(\vec{k},\,\omega)=\delta\mathcal{Z}(\vec{k},\,0)\,,
\eeqa
where
\beqa
\mathds{M}=
\label{c5}
\begin{bmatrix}
i\omega & -i k \rho_0 & 0 & 0 \\[4pt]
i k P_\rho & k^2 \nu -i\omega \rho_0 & i k P_T & i k P_\phi \\[4pt]
-i\omega P_\rho & 0 & i\omega \rho_0 C_P - i\omega P_T - \chi k^2 & -i\omega P_\phi \\[4pt]
\gamma P_\rho & i k \xi & \gamma P_T & -i\omega + \gamma P_\phi
\end{bmatrix}\,.
\eeqa
	
and matrices $\delta \mathcal{Z}(\vec{k},\,\omega)$ and $\delta \mathcal{Z}(\vec{k},\,0)$ are 
\beqa
		\delta \mathcal{Z}(\vec{k},\omega)=
		\begin{bmatrix}
			\delta \rho(\vec{k},\omega) \\
			\delta v(\vec{k},\omega)\\
			\delta T(\vec{k},\omega)\\
			\delta \phi(\vec{k},\omega)
		\end{bmatrix};
		\delta \mathcal{Z}(\vec{k},0)=
		\begin{bmatrix}
-\delta\rho(\mathbf{k}, 0) \\[4pt]
\rho_0 \, \delta v(\mathbf{k}, 0) \\[4pt]
P_\rho \, \delta\rho(\mathbf{k}, 0) + \left( P_T - \rho_0 C_P \right) \delta T(\mathbf{k}, 0) + P_\phi \, \delta\phi(\mathbf{k}, 0) \\
\delta\phi(\mathbf{k}, 0)\,.
\end{bmatrix}\,.
		\label{c6}
		\eeqa 
Using Eqs.\eqref{c5} and \eqref{c6}, and solving Eq.\eqref{c4}, the density fluctuation is obtained as:
\beqa
\label{eq23}
\delta {\rho}(\mathbf{k},\omega) 
&=& \left( -\mathbf{M}^{-1}_{11} + P_\rho \mathbf{M}^{-1}_{13} \right) \delta\rho(\mathbf{k}, 0) 
+ \rho_0 \mathbf{M}^{-1}_{12} \delta v(\mathbf{k}, 0) \nn\\
&&\quad + (P_T - \rho_0 C_P) \mathbf{M}^{-1}_{13} \delta T(\mathbf{k}, 0) 
+ \left( P_\phi \mathbf{M}^{-1}_{13} + \mathbf{M}^{-1}_{14} \right) \delta\phi(\mathbf{k}, 0)
\eeqa
Since, the correlation between two independent thermodynamic variables, say, $Q_i$ and $Q_j$  vanishes {\it{{\it{i.e.,}}}}
\beqa
\Big< \delta Q_{i}(\vec{k},\omega)\,\delta Q_{j}(\vec{k},0)\Big>=0, \,\,\,\, i\neq j\,.
\label{eq25}
\eeqa

The required correlator, $\mathcal{S^\prime}_{\rho\rho}(\vec{k},\omega)$ is obtained as:
\beqa
\mathcal{S^\prime}_{\rho \rho}(\vec{k},\omega)&=&\Big< \delta \rho(\vec{k},\omega)\,\delta \rho(\vec{k},0)\Big>\nn\\
&=&\Big[- \mathds{M}^{-1}_{11}+P_{\rho}\, \mathds{M}^{-1}_{13}\Big ]\, \Big< \delta \rho(\vec{k},0)\delta \rho(\vec{k},0)\Big>\,.
\eeqa
Finally, the $\Snn$ is defined as: 
\begin{equation}
\mathcal{S}_{\rho \rho}(\vec{k},\omega)=\frac{\mathcal{S^\prime}_{\rho \rho}(\vec{k},\omega)}
   {\Big< \delta \rho(\vec{k},0)\delta \rho(\vec{k},0)\Big>}\,.
\label{eq26}
\end{equation}	
Explicitly, it takes a form:
\beqa
\mathcal{S}_{\rho\rho}(\mathbf{k}, \omega) = \frac{\mathcal{N}_3 \omega^3 + \mathcal{N}_2 \omega^2 + \mathcal{N}_1 \omega + \mathcal{N}_0}{\mathcal{D}_4 \omega^4 + \mathcal{D}_3 \omega^3 + \mathcal{D}_2 \omega^2 + \mathcal{D}_1 \omega} \,,
\eeqa
where
\begin{subequations}
\label{Srhorho_power_coeffs}
\beqa
\mathcal{N}_3 &=& i \rho_0 (\rho_0 C_P - P_T), \\
\mathcal{N}_2 &=& -\left[ \rho_0 \chi k^2 + \gamma P_\phi \rho_0^2 C_P + k^2 \nu (\rho_0 C_P - P_T) \right], \\
\mathcal{N}_1 &=& - i k^2 \left[ \rho_0 \gamma P_\phi \chi + k^2 \nu \chi + \nu \gamma P_\phi \rho_0 C_P + \rho_0 C_P P_\phi \xi + \rho_0 P_\rho P_T \right], \\
\mathcal{N}_0 &=& \chi k^4 P_\phi \left( \gamma \nu +\xi \right), \\
\mathcal{D}_4 &=& \rho_0 (\rho_0 C_P - P_T), \\
\mathcal{D}_3 &=& i \left[ \rho_0 \chi k^2 + \rho_0^2 \gamma P_\phi C_P + k^2 \nu (\rho_0 C_P - P_T) \right], \\
\mathcal{D}_2 &=& - \rho_0 \gamma P_\phi \chi k^2 - k^4 \nu \chi - k^4 \nu \gamma P_\phi \rho_0 C_P - k^2 P_T P_\phi  - i k^2 P_T P_\phi \xi \nn\\
&&+ i k \gamma P_\phi P_T (\rho_0 C_P - P_T) - k^2 \rho_0 P_\rho (\rho_0 C_P - P_T) - i k^2 \rho_0 P_\rho P_T  \nn\\
&&+ k^2 \rho_0 P_\rho P_\phi \gamma P_T \xi, \\
\mathcal{D}_1 &=& - i k^4 \nu \gamma P_\phi \chi + k^2 \gamma P_T P_\phi^2 + k^3 P_\phi P_T\xi - k^3 \gamma P_\phi P_T \chi \nn\\
&&- i k^4 \rho_0 P_\rho \left( \chi + \rho_0 \gamma P_\phi C_P \right) + i k^2 \rho_0 \gamma P_\rho P_T P_\phi - i k^2 \rho_0 \gamma P_\rho P_\phi (2P_T - \rho_0 C_P)\,.
\eeqa
\end{subequations}

\section{The transport coefficients and the thermodynamic response functions}
\label{apD}
The thermodynamic response functions and transport coefficients used in our analysis are summarized in Table~\ref{tab:scaling_laws}, following the critical scaling relations discussed in Ref.~\cite{Kadanoff}.

\begin{table}[htbp]
\centering
\setlength{\tabcolsep}{30pt}
\renewcommand{\arraystretch}{0.9}

\begin{tabular}{l l l}
\hline\hline
\textbf{Quantity} & \textbf{Regime} & \textbf{Scaling law} \\
\hline

$C_P$ &
All frequencies  &
$C_P \sim t^{-\gamma}$ \\

\hline

$C_V$ &
All frequencies &
$C_V \sim t^{-\alpha/2}$ \\

\hline

Thermal conductivity ($\chi$) &
Low frequency &
$\chi \sim t^{-\gamma+\nu}$ \\

&
High frequency &
$\chi \sim t^{2\nu-\gamma-\alpha/2}$ \\

\hline

Shear viscosity ($\eta$) &
Low frequency &
$\eta \sim t^{\nu-\gamma}$ \\

&
High frequency &
$\eta \sim t^{2\nu-\gamma+\alpha/2}$ \\

\hline

Bulk viscosity ($\zeta$) &
Low frequency &
$\zeta \sim t^{-\gamma-\nu+3\alpha/2}$ \\

&
High frequency &
$\zeta \sim t^{2\nu-2+\alpha/2}$ \\

\hline\hline
\end{tabular}
\caption{Critical scaling behaviors of thermodynamic and transport
quantities near the critical point. Here,
$t=(T-T_c)/T_c$, with $\nu\simeq 2/3$, $\gamma\simeq 4/3$,
and $\alpha\simeq 0.1$.}
\label{tab:scaling_laws}
\end{table}

\bibliography{NS}

\end{document}